\documentclass[9]{acm_proc_article-sp}
\usepackage{url}
\usepackage{float}
\usepackage{setspace}

\begin{document}

 \title{Semantic Identification Attacks on Web Browsing}

\numberofauthors{1} 
\author{
\alignauthor
Neel Guha \\
\affaddr{Stanford University}\\
\affaddr{CA, USA}\\
\email{nguha@stanford.edu}
}

\maketitle
\begin{abstract}

We introduce a \textit{Semantic Identification Attack}, in which an adversary uses semantic signals about the pages visited in one browsing session to identify other browsing sessions launched by the same user. This attack allows an adversary to determine if two browsing sessions originate from the same user regardless of any measures taken by the user  to disguise their browser or network. We use the MSNBC Anonymous Browsing data set, which contains a large set of user visits (labeled by category) to implement such an attack and show that even very coarse semantic information is enough to identify users. We discuss potential countermeasures users can take to defend against this attack. 

\end{abstract}

\section{Introduction}
Online privacy is becoming an increasingly important issue for users, policy makers, and academic researchers. In recent years there has been significant work highlighting the fragility of user privacy by exposing threats ranging from the de-anonymization of public data sets to third party tracking in online browsing. 

In particular, an area of focus has been the ability for advertisers to identify users across browsing sessions. Many privacy experts advocate that users should regularly clear browsing cookies to prevent third party trackers from building detailed profiles and linking their webpage visits.  Other experts recommend private browsing modes (such as incognito mode on Google Chrome) to prevent the browser from storing any persistent browsing data and to ensure that web cookies are never reused. There are even more powerful tools, such as Tor or Brave, a new open source browser focused explicitly on blocking advertisements and third party trackers.  

Unfortunately, these measures are not adequate. In this paper, we introduce a \textit{semantic identification attack}, a new attack that can  identify users across multiple browsing sessions.  In this attack, an adversary can, in some cases, leverage semantic signals - features derived from the content viewed by a user (words on pages viewed, page titles, links in/out of page, etc) -  to identify when two browsing sessions have been launched by the same user. If one of the sessions contains personally identifying information, the adversary now has a way of linking this information to other sessions launched by the user.

A user is unlikely to visit the same urls in every session. However, the urls in a given user's browsing sessions are unlikely to be drawn completely at random from all the urls on the web. If a user's web browsing is characterized by a set of tasks (checking certain news topics, reading specific blogs, etc), their browsing sessions will consistently contain visits to pages/websites relevant to these tasks. The distribution of these visits can be viewed as a user specific session fingerprint. Though a user's browsing sessions are likely to deviate and contain other behavior as well, the presence of these pages fingerprints the user to the session. 

The ability to extract a browsing fingerprint and use it to link together a user's sessions has significant implications for online privacy. An adversary that could identify the browsing fingerprint for a user could determine with high probability when a certain  browsing session belongs to that user even when they use multiple devices, regularly clear cookies or obfuscate their identity through proxies.

While there has been prior work in identifying users across browsing sessions, nearly all methods have relied on the browser or network metadata for identifying users. Network signals and browser metadata signatures can be spoofed or disguised through extensions and proxies. However, semantic data about browsing sessions will still be available to websites and even to third parties such as advertisers. We demonstrate that this kind of data can be used to identify and track users. 

In this paper, we first review past work and highlight the differences between prior approaches and ours. We then explain the intuition behind the semantic identification attack and present several different strategies. Finally, we evaluate the attacks on a collection of user browsing data and discuss the countermeasures.

\section{Related Work}

There has been significant work on fingerprinting users on the basis of their browser\cite{Yen2009}\cite{mowery2011fingerprinting}. \cite{Eckersley2010}  and \cite{host-fingerprinting-and-tracking-on-the-webprivacy-and-security-implications} describe how users can be uniquely identified by analyzing system fonts, screen resolution, information stored in HTTP headers (ie, HTTP User Agent strings) and coarse IP prefix information (users can test their fingerprints at \cite{pano} or \cite{amiunique}). Additionally, \cite{MS12} showed that it is possible to fingerprint users by leveraging the subtle differences in the ways different browsers render the same text. These differences provide consistent, unique fingerprints that  can be used to track users.

However, there has been little work on identifying user web browsing without using browser metadata or network information as features. While both have been used with large success to fingerprint users, one could argue that more secure browsers or more privacy awareness could diminish the success they provide adversaries. Our paper shows however, that even if a user were take these precautions, an adversary could still identify them on the basis of the content of their browsing. Browsing history is a strong predictor for learning user attributes, suggesting that semantic signals from browsing sessions could convey significant discriminating information \cite{goel2012does}.  Jones et al \cite{Jones:2007:IKY:1321440.1321573} show that it is possible to infer personal information (gender, age, location, etc) from query logs and link a particular user to a query stream when provided with some background information on that user.  \cite{Kumar:2007:AQL:1242572.1242657} demonstrates that even a countermeasure like token-based hashing of queries fails to prevent an adversary from identifying users. \cite{5708146} proposes  the ZEALOUS algorithm, which is capable of achieving strong privacy guarantees on query logs.  \cite{Novak:2004:AW:988672.988678} focuses on determining when a single individual is masquerading behind multiple aliases on a web forum using the  content present in forum posts. it is similar to our work in that it aims to use semantic information to uniquely identify individuals. However, \cite{Novak:2004:AW:988672.988678} is applied to a forum (as opposed to web browsing) and leverages significantly richer data. 

Another closely related field of study is differential privacy\cite{narayanan2010myths}. Though differential privacy is primarily concerned with the privacy guarantees of statistical databases and their individual entries,  some of the core concepts and ideas are applicable in our case. \cite{Dwork2008} describes some key findings. \cite{agrawal2000privacy} also present techniques for preserving privacy in data mining applications. The 2006 AOL scandal demonstrated that there are techniques available to identify users from anonymized web search data \cite{aol_scandal}. Narayanan and Shmatikov presented a further class of de-anonymization techniques by using the IMDB dataset to identify individual users in the Netflix Prize Dataset \cite{Narayanan:2008:RDL:1397759.1398064}.

\section{Browsing Fingerprints}
We first describe browsing fingerprints and the intuition behind semantic identification attacks. 

\subsection{Intuition}

A semantic identification attack leverages the fact that web browsing tends to be a highly personal activity. Let  $D_u$ be the  distribution over all webpages of the likelihood that a user $u$ visits a webpage. Thus, for a given page $p$ and a user $u$, $D_u(p)$ is the probability that user $u$ visits web page $p$.  $D_u$ is not a uniform distribution - it is obvious that there are certain pages $p_i$  that a user is more likely to visit and that $D_u$ is skewed towards these pages. Even though a user is unlikely to repeatedly visit the same web page, they are like to repeatedly visit similar types of webpages. A user is more likely to visit pages that pertain to the tasks, interests, and habits motivating the user's browsing.

The intuition behind a semantic identification attack is that many users have a unique $D_u$ - the distribution of pages they're likely to visit is shared with few others, if any. We thus refer to $D_u$ as the user's browsing fingerprint - a signature unique to the user, that can be used to identify them. For a given session $s_i$ launched by a user $u_i$, the goal of a semantic identification attack is to extract the browsing fingerprint of $s_i$  (some representation of $D_i$).  Given two sessions $s_i$  and $s_j$, the adversary can then compare the browsing fingerprints (the extracted representations of $D_i$ and $D_j$). Because browsing fingerprints are unique to users, an adversary can deduce that $s_i$ and $s_j$ were launched by the same user if their browsing fingerprints are similar enough. 

The likelihood a user visits a page is dependent on the content of that page.  In order to extract a browsing fingerprint and some representation of $D_u$, an adversary must have some knowledge of the topics, entities or words on each page visited by the user. We refer to these as "semantic signals". At their core, semantic signals are features which capture the interests, habits, and preferences of the user. Across multiple sessions, the pages a user visits are likely to share some semantic similarity. If we treat each page as a set of semantic  signals, then $D_u$ can be thought of a user specific distribution over these semantic signals. The browsing fingerprint for a user is a representation of this distribution.

The following example illustrates this. Imagine a computer science student from California studying in Cambridge, MA. This user's browsing interests consist of their favorite sports teams (the Golden State Warriors), local Boston news, and Stack Overflow. The browsing fingerprint for this user (and their respective $D_u$) would likely indicate a skew towards pages covering these topics. This user is characterized by their interests - the locality of their news, specific sports personalities, and their professional interests. Individually, none of these interests are unique and are shared by hundreds of thousands of individuals. When combined however, the set of users who lie at this intersection is much smaller. 

If an adversary saw two browsing sessions which captured these interests - the Golden State Warriors, Boston news and Stack Overflow, the adversary might predict that both sessions originated from the same user as their browsing fingerprints were similar. If the user's interests included something esoteric (like Vietnamese vegetarian cooking), it becomes significantly easier for the adversary to extract a browsing fingerprint (assuming Vietnamese vegetarian cooking is a more unique and hence a distinguishing interest than the other topics). 

Let us explore another simple example. Suppose an adversary identifies from a user's browsing session that they're  interested in privacy research. This information alone is unlikely to uniquely identify or distinguish this user as plenty of individuals are interested in privacy research and spend time reading about it online. However, suppose that the adversary was also able to identify that this user was also interested in Loreena Mckennitt's music. Loreena Mckennitt is a famous musician, and likely to have a wide online following. Individually, neither of these interests are particularly unique or distinguishing - a browsing fingerprint encapsulating either of them would be relatively useless. However, the intersection of those who both listen to her music and are interested in privacy research is likely to be rather small. A browsing fingerprint that captures this intersection is significantly more powerful and unique. Whenever the adversary sees a browsing session that involves privacy research and Loreena McKennitt, the adversary will with strong probability predict that it was launched by the same user. In this manner, the adversary may be able to track the user across multiple sessions.

In this work we focus on determining whether two sessions originate from the same user (and thus share the same fingerprint). We do this by extracting browsing fingerprints for each session and comparing those fingerprints to determine whether they originate from the same user. 

For any given user, there are several cases for which a browsing fingerprint will be difficult to extract/leverage.
\begin{enumerate}
\item Individuals who do not browse frequently or browse in a very general manner (visiting only the homepage of certain websites) are unlikely to have a distinctive browsing fingerprint. Though we may be able to learn their fingerprint, it is not unique enough to identify them in other browsing sessions. Thus, any attack will succeed at identifying "outlying" users - users who's behavior is sufficiently unique that it can be identified. 
\item Though we assume a user's browsing will largely follow $D_u$, there are likely to be browsing sessions which constitute atypical behavior and cannot be identified. Imagine a user who is going on a vacation (not a frequent occurrence) and decides to use the web to plan their trip. If their browsing session consists exclusively of trip planning, it is unlikely to correlate with any of their prior sessions and will thus go unidentified. Similarly, a user who was browsing in response to an emergency (like an earthquake) is unlikely to repeat their browsing often.
\end{enumerate}

There are several statistical tests for quantifying the difference between two binned distributions. Most of them - like the chi-squared test - assume some prior knowledge about the distribution and the expected values of each bin. Since an adversary can make no assumptions about the $D_u$ , these tests are not applicable here. We thus introduce our own methods later in the paper.  

\subsection{Semantic Signals}

Semantic signals capture the content of each page at some level of granularity. They allow the adversary to extract the user based browsing fingerprint and build some approximation of $D_u$. Some examples of semantic signals include:

\begin{itemize}
\item The words on the page visited
\item Categorical classifications for the page visited ("news", "social media", "cooking")
\item The title of the page
\end{itemize}

The success of a semantic identification attack is highly dependent on the granularity of the semantic signals. Rich signals - such as a listing of every word on every page visit will allow the adversary to extract a more powerful browsing fingerprint. Weak signals - such as a simple classification of each webpage (ie "news"), is rather poor and will lead to fewer unique browsing fingerprints. 

The semantic signals available to an adversary depend on the adversary's capabilities and the position from which they launch the attack. In this work, we demonstrate that even a weak adversary with poor signals can still uniquely identify some users. 

\subsection{Fingerprint Granularity}
Given a browsing session, we can either extract a single fingerprint for the entire session (a cross site fingerprint) or multiple fingerprints, one for each website in that session.  Most web browsing sessions consist of page visits to multiple different websites. If we extract a cross site fingerprint, then we extract a single fingerprint that describes the user's browsing behavior over all of these websites. If we extract site fingerprints, then for each website in the session we extract a single fingerprint describing the user's behavior on that website in that session. Because these fingerprints are being extracted on a per site basis, we extract multiple fingerprints per session (one for each of the websites visited). As each of these fingerprints reflect more localized behavior (on a single website), they are less likely to be unique across users. However, we can use the aggregation of fingerprints across websites to identify users.

Though individual website based fingerprints are less insightful, they are more resistant to user defenses. A cross site fingerprint for example, may be hard to collect if the user blocks third party cookies (preventing cross site  data from being collected by an adversary). However, the website a user visits will always know the pages clicked on by the user. If this website can extract a granular site specific fingerprint for the user, they can use that to identify the user when they revisit the website. Thus, even if a user uses incognito mode to make these visits, the website may be still be able to track them.

There are relatively few publicly available browsing data sets making validation of any of these methods significantly challenging.  Due to constraints with available data for validation, in this paper we describe our methods in terms of website specific fingerprints. We now discuss several of these methods. 

We assume that an adversary has access to the semantic signals for a set of $S$ browsing sessions originating from a set of $U$ different users. We assume that every $s \in S$ was launched by exactly one $u \in U$ but that most $u$ launched multiple sessions in $S$.

\subsection{Assumptions}
We make the following assumptions on user browsing: 
\begin{enumerate}
\item A semantic identification attack assumes that for a user $u$, $D_u$ is skewed towards certain pages. If $D_u$ was a uniform distribution, and all pages had an equal likelihood of being visited and browsing sessions would resemble a random sampling of the web. In such a case, it would be difficult to extract a unique browsing fingerprint.  Because we assume $D_u$ is skewed (and nonuniform), we know there are some pages a user is more likely to visit and we're able to extract browsing fingerprints that are specific to a user.
\item We have some semantic information about the activity of a user in a browsing session. This could comprise of the words on every page visited by the user, categorical labels for each page, or a set of entities capturing the range of topics spanned by the session. These are the features leveraged by the adversary in order to identify users.
\item Conversely, we assume that the adversary has access to no other information about the browsing session. These could include geographic location, device identity, network addresses, or browser metadata signatures. Later in the paper we discuss how access to this kind of data could improve a semantic identification attack.  
\end{enumerate}

\section{Attack Strategies}

Given two sessions and their associated semantic features, we wish to determine whether or not they originated from the same user. In order to do so we must extract some representation of the browsing fingerprints and calculate some similarity score between them. We present two approaches for making this determination. First however, we briefly discuss the dataset we use to validate these methods. 

\subsection{Dataset}
The lack of datasets of real browsing sessions  severely constrains our ability to test different attack strategies. Before we present our strategies, we briefly describe the dataset used to validate them. 

We used the MSNBC Anonymous Browsing Data set (\cite{msnbc_data}). The data contains a list of pages visited on msnbc.com for 990,000 users over a 24 hour period (on September 28, 1999). Each page is represented as a page category (a page can be "news", "homepage", "local", "tech", etc). The granularity of our data is restricted to the page category, for a given page that a user visited, we only know what category the page falls under. We have no way other information about that particular visit. Below is a sample representation of the data: \[
\begin{bmatrix}
     \{1,3,5,1,3,4,4\} \\
    \{1,3,7\} \\
    \{5,7,9,14,15\} 
\end{bmatrix} \]
Each line in the matrix above corresponds to the browsing activity for a single user. For example, in this case the data tells us that user corresponding to the second line visited a page of category type 1, followed by category type 3, and finally category type 7. This user visited no other pages on msnbc.com during this time period. 

These categories are the "semantic signals" for each of the user's page visits. Despite the coarse grained information(very different pages could be labelled with the same category), we show it is still possible to construct unique fingerprints for a subset of users and to correctly identify sessions originating from them.

\subsubsection{Fingerprint Representation}
We represent each session as a vector of the proportions of each of the 17 categories of pages visited in each session. Thus if a session contained the visits to the following pages: $$[1,2,2,2,2,3,4,3,3,3]$$
The corresponding vector representation would be $$[0.1, 0.4, 0.4, 0.1, 0,0,0,0,0,0,0,0,0,0,0,0,0]$$
Since visits to pages of category "1" represent 10\% of the session's page visits, pages of category "2" represent 40\% of the session's page visits, and etc. Each session can be seen as a unit vector in a 17 dimensional space, where each dimension corresponds to a category. 

This representation is general and intuitive, but does fail to capture some potentially interesting information such as the order in which the user looks at pages in different categories.

\subsection{Pairwise Similarity}

In this approach, we compute a similarity score for every single pair of sessions. If the score for two sessions exceeds a threshold, then we predict that two sessions originate from the same user. For two sessions represented as the vectors $s_i$ and $s_j$, we calculate $Score(s_i,s_j)$ in the following manner. 
$$ Score(s_i, s_j) = \dfrac{Sim(s_i,s_j)}{\sum_{k=0}\dfrac{1}{Sim(s_i,s_k)} + {\sum_{k=0}\dfrac{1}{Sim(s_j,s_k)}}} $$where $Sim$ is some similarity metric. Intuitively, this score can be thought of as the combination of a similarity score (between the two sessions) and a dissimilarity score. We discuss both in more detail below. 

\subsubsection{Similarity Score}
For calculating the similarity between two sessions $s_i$ and $s_j$, we use the cosine similarity metric, a common measure in information retrieval. Given two vectors in an n-dimensional space, cosine similarity measures their distance as a function of the angle between them. 
$$ Cosine (s_i, s_j) = \dfrac{s_i \cdot s_j}{\vert \vert s_i \vert \vert \vert \vert s_j \vert \vert} $$
\subsubsection{Dissimilarity Score}
Simply measuring whether two sessions are similar is not sufficient enough to capture whether or not they may originate from the same user. Our underlying goal is to be able to capture and compare the browsing fingerprints of two sessions. Thus, we need to weight the similarity of two sessions by how dissimilar those sessions are from all other sessions. If we have two sessions $s_i$ and $s_j$  such that $s_i$ is similar to $s_j$ but both $s_i$ and $s_j$ are similar to the bulk of the sessions in our data set, we are less confidant that $s_i$ and $s_j$ originate from the same user. it is probable that $s_i$ and $s_j$  (and the sessions they're similar too) belong to a mass of  users whose behavior is too shallow or generic to discern. Conversely, if $s_i$ and $s_j$ were similar to each other but different from other sessions, we'd be significantly more confidant that both sessions originated from the same user. 

\subsection{Supervised Learning}

Alternatively, as opposed to devising our own similarity score, we can rely on machine learning to learn some scoring function. We rely on a multi-layer perceptron classifier from the scikit-learn python machine learning library\cite{scikit}.  For two sessions $s_i$ and $s_j$  with the category proportion vectors $c_i$ and $c_j$, our classifier takes as input the $\vert c_i - c_j \vert$  (the absolute value of the difference between the category proportions). We demonstrate that even with such a simple classifier, we're able to achieve success with our model. 

\subsection{Metrics}
We now define the metrics used to evaluate the success of a semantic identification attack. We consider our attack in the following framework. Assume we have a set of browsing sessions $S = \{s_1,s_2,..,s_n\}$ where each $s_i$ was launched by a single user and there is at least one other $s_j \in S$ such that $s_j \neq s_i$ and both $s_j$ and $s_i$ were launched by the same user. For every pair of sessions in $S$ the goal for the adversary is to identify whether they share the same user. 
We can measure an adversary's success along two dimensions. 

\subsubsection{Precision}
Precision is defined as the proportion of session pairings identified by the adversary that are correct. Let $S_p$ be the set of pairs of sessions such that for each pair the adversary has predicted both sessions originate from the same user. Let $S_t$ be the set containing all pairs of sessions in $S$ that originate from the same user. We can then formally define:
$$\text{Accuracy} =  \dfrac{S_t \cap S_p}{S_p}$$
\subsubsection{Recall}
We define recall as the proportion of same origin session pairs successfully identified by the adversary. More formally: 

$$\text{Recall} = \dfrac{S_t \cap S_p}{S_t} $$ 

\subsubsection{Reach}
We define reach as the number of unique users for whom the advertiser has successfully paired at least two sessions. Reach captures the "user footprint" of the semantic identification attack and allows to quantify the number of users (not sessions) that an adversary has successfully compromised. 

\subsubsection{Success}
it is important to note that in this framework, there are many possible definitions for a "successful" attack. In the ideal case, an adversary would prefer an attack strategy which grantees high accuracy and high reach - identifying sessions from a lot of users with  high probability of correctness. However depending on an adversary's goals, different dimensions may be prioritized. An advertiser for example, seeking to broadly identify users in order to show related advertisements, may care less about accuracy and more about reach. Since ads are generally viewed as terrible, they may care more about maximizing the number of users they reach (as opposed to the proportion of time they're correct). 

Conversely, a different adversary may care more about maximizing their accuracy (as opposed to their reach). If they're trying to closely track a set of users, they may care more about the strength of their predictions and less about accuracy.

\section{Results}
We now outline the results achieved by applying these techniques to the MSNBC dataset described earlier.

\subsection{Data Characteristics}
The majority of users visit only a handful of pages before leaving msnbc.com. Most of these page visits are to "news",  "on-air" or "sports" pages (table \ref{page_cat_dist}).

\begin{table}[h]
\begin{tabular}{|l|l|l|}
\hline
Page Category & Frequency & Proportion \\
\hline
front page & 940469& 0.200 \\
news &452387 & 0.096 \\
tech &207479 &0.044 \\
local &386217 &0.082 \\
opinion &151409& 0.032 \\
on-air &414928& 0.088 \\
misc& 305615& 0.065 \\
weather &439398& 0.093 \\
health &196614& 0.041 \\
living &131760& 0.028 \\
business &96817& 0.020 \\
sports &264899& 0.056 \\
summary& 216125& 0.045 \\
bulletin board service &395880& 0.084 \\
travel &56576& 0.012 \\
msn-news  &25249& 0.005 \\
msn-sports &16972& 0.003 \\
\hline
\end{tabular}
\caption{Distribution of page views over page categories}
\label{page_cat_dist}
\end{table}

\subsection{Session Creation}
Unfortunately, because there are no time stamps for page visits, it is impossible to determine from this data when a user's session begins and ends. Even if a user launched multiple sessions on msnbc.com over the 24 hour window, all page visits have been aggregated onto a single line. If clear session demarcations had been provided, then we could have used them to test our attacks by seeing if our methods could identify sessions from the same user. Since we do not have these demarcations however, we have to artificially create our own sessions. We introduce two methods to split a single user's browsing into multiple discrete sessions. We describe both approaches below.

\subsubsection{Homepage Partition}
We utilize the intuition that any user starting a new session on msnbc.com will  start by visiting the homepage. Thus for a given user who visits the list of page $ [p_1,p_2,..p_n]$, we partition this list into sublists whenever a page $p_j $  is the home page. For example, if a user visited 50 pages, and their 20th page visits and 45th page visits were to the homepage, then we would partition their browsing into 3 separate sessions $[[p_1,...,p_{19}],[p_21,...,p_{44}],[p_{46},...,p_{50}]]$ . In order to ensure that sessions are long enough to extract some sort of browsing fingerprint from, we also discard all sessions whose length is below some minimum threshold $k$.  

\subsubsection{Random Partition}
In the random partition strategy, we fix the session length at some $k$ and randomly select $k$ pages at a time from the user's browsing list (without replacement) until the pages left is fewer than $k$. Each of the $k$ samples corresponds to a single session from that user. For example, if a user visited 50 pages and $k=20$, we would randomly select 20 out of the 50 pages as the first session, 20 out of the remaining 30 pages as the second session, and discard the remaining 10 pages (since $10 < k$). Thus, for this user we would have artificially created 2 sessions. 

While visiting a homepage may signal a user starting a new session, it could also indicate a context switch within a session. A user who was previously browsing the news may go back to the homepage in order to start browsing sports. Partitioning on the homepage would split these into two separate sessions - each with very different characteristics. As described later in the paper, this is likely the case in our results. Table \ref{rp_stats} and \ref{hp_stats} describe the number of sessions generated (and from how many users) for random and homepage partitioning respectively. 

\begin{table}[h]
\begin{tabular}{|l|l|l|}
\hline
 Session Length ($k$)&	Number of Users&	Number of Sessions \\
\hline
10	&35368	&102082 \\
15	&12777	&38731\\
20	&5888	&19403\\
25	&3128	&11439\\
30	&1891	&7609\\
35	&1294	&5616\\
\hline
\end{tabular}
\caption{Session frequencies and user counts with different $k$ using random partition}
\label{rp_stats}
\end{table}
\vspace{-1mm}
\begin{table}[h]
\begin{tabular}{|l|l|l|}
\hline
 Session Length ($k$)&	Number of Users&	Number of Sessions \\
\hline
10&2645&	6164 \\
15	&854&	2036 \\
20	&395&	973 \\
25	&242&	624 \\
30	&149&	396 \\
\hline
\end{tabular}
\caption{Session frequencies and user counts with different $k$ using homepage partition}
\label{hp_stats}
\end{table}

\subsection{Adversary Performance}
We now investigate the performance of our attack strategies on this sample data. 
We tested the attacks on both homepage partitioned and randomly partitioned sessions. For homepage partitioning we enforced a minimum session length of 35 pages (discounting all sessions with fewer than 20 pages from our analysis set). For randomly partitioned we fixed the session size at 35 pages.

\vspace{-6mm}
\subsubsection{Baseline }
We establish a baseline for our algorithms by randomly assigning scores between $[0,1]$  for every pair of sessions. We then determine the precision/recall scores as we adjust this cutoff. 
\vspace{-4mm}
\subsubsection{Random Partition}
We ran the attacks on sessions from random samples of 300, 500, 750 and 1000 users. Taking random samples allows us to simulate the variance in users an adversary will encounter when they launch an attack. The browsing fingerprint of a user is  valuable insofar as it is unique and distinguishable from the browsing fingerprints of other users. Thus, an adversary's ability to identify a specific user is highly dependent on the other users the adversary sees. Samples of different sizes provide us with an approximation of the attack's success on different sized websites. 
 
We ran 25 trials and ran the attacks on samples of 300, 500, 750 and 1000 users during each trial. For each sample size, we report both the average precision/recall/reach across all trials. We also report the precision/recall/reach of the trial in which the adversary was most successful at identifying users. From a privacy perspective, we care about the worst case attack is as significant as the average case. it is important to know the upper bound of the potential risk. 

The neural network was trained on the sessions for 100 users with 1 hidden layer and 100 neurons. Tables \ref{300_best_table},\ref{500_best_table},\ref{750_best_table}, and \ref{1000_best_table} contains the results of the worst case attack on the samples of 300, 500, 750 and 1000 users respectively. The precision, recall and reach values for each of the attacks correspond to the values that maximized the F1 score. Table \ref{300_avg_table}, \ref{500_avg_table}, \ref{750_avg_table}, \ref{1000_avg_table}, correspond to the average of 300, 500, 750 and 1000 user samples (respectively) across all 25 trials. The values in the table are those that maximize the F1 score. In nearly every sample, both the average and best case performance significantly exceeded the baseline. However, as the size of the user sample grew, the performance of the attack appeared to decline.  

\begin{figure}[h]
\includegraphics[scale=0.4]{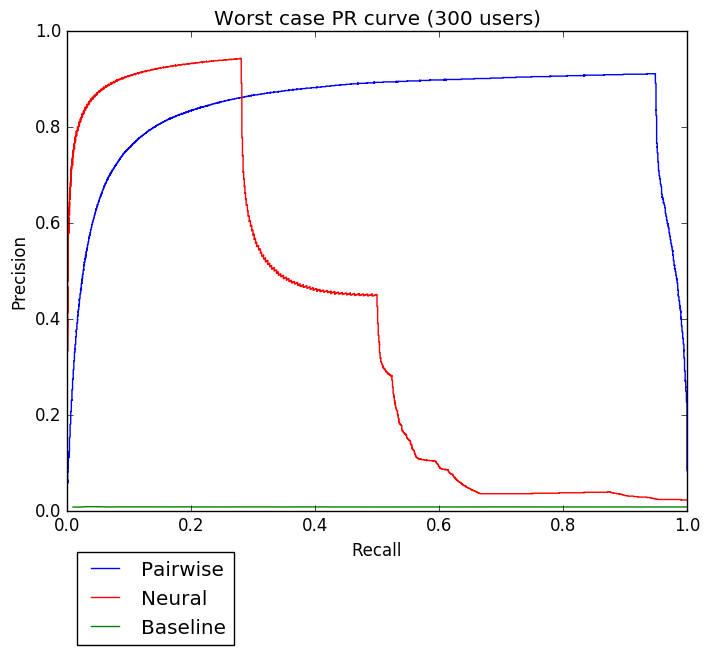}
\caption{Recall vs precision curve for worst case attack on 300 user sample}
\label{pr_300_best}
\end{figure}
\vspace{-3mm}
\begin{table}[H]
    \begin{tabular}{|l|l|l|l|l|}
    \hline
    Attack    & F1 & Precision & Recall   & Reach \\ \hline
    Pairwise   & 0.464806& 0.911000  & 0.949000 & 45    \\
    Neural Net &0.237119 & 0.451000  & 0.500000 & 5     \\
    Baseline   & 0.008431 &0.008532  & 0.713416 & 227   \\ \hline
    \end{tabular}
    \caption{Worst case attack on 300 user sample}
    \label{300_best_table}
\end{table}
\vspace{-2mm}
\begin{figure}
\includegraphics[scale=0.4]{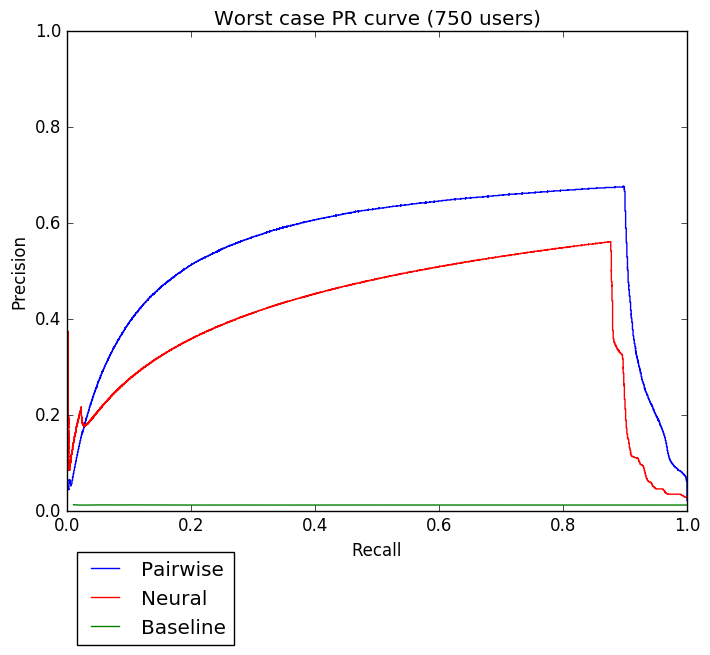}
\caption{Recall vs precision curve for worst case attack on 750 user sample}
\label{pr_750_best}
\end{figure}
\vspace{-3mm}
\begin{table}[H]
    \begin{tabular}{@{}|l|l|l|l|l|@{}}
    \hline
    Attack     &F1 & Precision & Recall   & Reach      \\ \hline
    Pairwise   & 0.274883 &0.392010  & 0.920000 & 145 \\
    Neural Net & 0.168720 &0.424528  & 0.280000 & 42  \\
    Baseline   & 0.008431 &0.008532  & 0.713416 & 227        \\ \hline
    \end{tabular}
    \caption{Average performances on 300 user sample}
    \label{300_avg_table}
\end{table}
\vspace{-4mm}
\begin{table}[H]
    \begin{tabular}{@{}|l|l|l|l|l|@{}}
    \hline
    Attack     & F1 &Precision & Recall   & Reach \\ \hline
    Pairwise   & 0.422236 &0.787000  & 0.911000 & 74    \\
    Neural Net & 0.438835 &0.861000  & 0.895000 & 13    \\
    Baseline   & 0.015561 &0.015906  & 0.717323 & 392   \\ \hline
    \end{tabular}
     \caption{Worst case attack on 500 user sample}
         \label{500_best_table}
\end{table}
\vspace{-4mm}
\begin{table}[H]
    \begin{tabular}{|l|l|l|l|l|}
    \hline
    Attack     & F1 &Precision & Recall   & Reach      \\ \hline
    Pairwise   & 0.326394 &0.508936  & 0.910000 & 169 \\
    Neural Net & 0.282179 &0.413180  & 0.890000 & 136 \\
    Baseline   & 0.015561 &0.015906  & 0.717323 & 392        \\ \hline
    \end{tabular}
    \caption{Average performance on 500 user sample}
    \label{500_avg_table}
\end{table}
\vspace{-4mm}
\begin{table}[H]
    \begin{tabular}{|l|l|l|l|l|}
    \hline
    Attack     & F1 & Precision & Recall   & Reach \\ \hline
    Pairwise   & 0.385857 &0.676000  & 0.899000 & 105   \\
    Neural Net & 0.342140 &0.561000  & 0.877000 & 28    \\
    Baseline   & 0.012289 & 0.012442  & 1.000000 & 750   \\  \hline
    \end{tabular}
    \caption{Worst case attack on 750 user sample}
    \label{750_best_table}
\end{table}
\vspace{-4mm}
\begin{table}[H]
    \begin{tabular}{|l|l|l|l|l|}
    \hline
    Attack     & F1 &Precision & Recall   & Reach      \\ \hline
    Pairwise   & 0.272032 &0.440999  & 0.710000 & 199 \\
    Neural Net & 0.201456 &0.284529  & 0.690000 & 192 \\
    Baseline   & 0.012289 &0.012442  & 1.000000 & 750        \\ \hline
    \end{tabular}
    \caption{Average performances on 750 user sample}
        \label{750_avg_table}
\end{table}
\vspace{-4mm}
\begin{table}[H]
    \begin{tabular}{|l|l|l|l|l|}
    \hline
    Attack     & F1 &Precision & Recall   & Reach \\ \hline
    Pairwise   & 0.336260 &0.578000  & 0.804000 & 154   \\
    Neural Net & 0.384924 &0.946000  & 0.649000 & 5     \\
    Baseline   & 0.015088 &0.015326  & 0.970913 & 955   \\ \hline
    \end{tabular}
    \caption{Worst case attack on 1000 user sample}
        \label{1000_best_table}
\end{table}
\vspace{-3mm}
\begin{table}[H]
    \begin{tabular}{|l|l|l|l|l|}
    \hline
    Attack     & F1 &Precision & Recall   & Reach      \\ \hline
    Pairwise   & 0.260535 &0.473010  & 0.580000 & 185 \\
    Neural Net & 0.254350 &0.417863  & 0.650000 & 211 \\
    Baseline   & 0.015088 &0.015326  & 0.970913 & 955        \\ \hline
    \end{tabular}
    \caption{Average performance on 1000 user sample}
        \label{1000_avg_table}
\end{table}

\subsubsection{Homepage partition}
\begin{figure}
\includegraphics[scale=0.4]{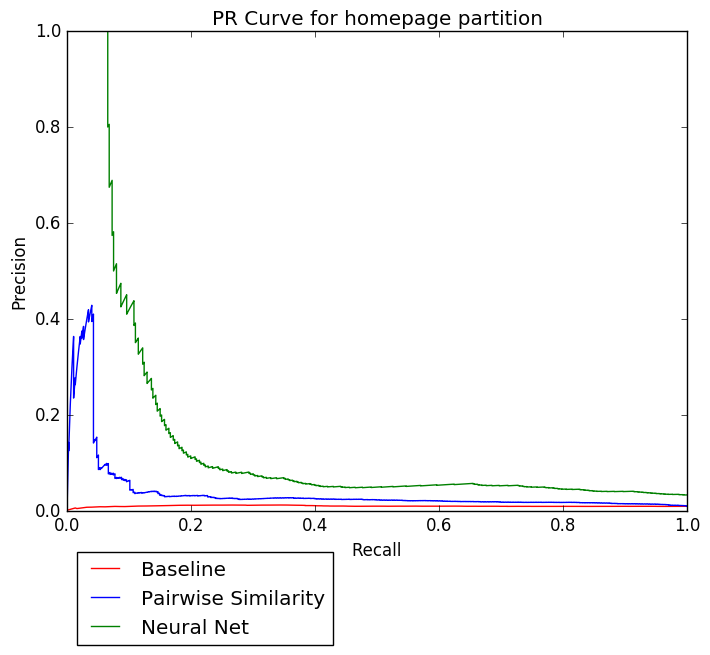}
\caption{Recall vs precision for homepage partitioned sessions with minimum 35 pages per session}
\label{35_pr_hp}
\end{figure}

Conducting homepage partitioning with a minimum session length of 35 pages resulted in 278 sessions from 100 users. Because of the limited size of the data, we ran our attack on the complete set (no sampling). Our neural network was trained on 50 users and contained 1 hidden layer with 100 neurons. Figure \ref{35_pr_hp} contains the PR curves for the different attacks. The pairwise method achieves a peak F1 score when precision is 0.098 and recall is 0.67 (reach is 3). The neural network achieves a peak F1 score when precision is 0.276  and recall is 0.136 (reach is 5). 

At lower recall values both attacks are successful at identifying users. However, both perform significantly worse on homepage partitioned sessions than on randomly partitioned sessions. 
\vspace{-1mm}
\vspace{-1mm}
\section{Defense Mechanisms}

We present a simple countermeasure in order to aid users in protecting privacy and defending against semantic identification attacks. For each session, we append $p$ pages where each page corresponds to a randomly selected category. Intuitively, these additional randomly distributed pages should add noise to the session and lessen the skew of $D_u$, making it harder for the attack to successfully extract the browsing fingerprint.  

We experimented with $p = {\{5,10,15\}}$  for a sample of 300 users and a session length of 35 pages. We ran 25 trials for each value of $p$  and we report the precision, recall, and reach for the worst case attacks for each value of $p$. Tables \ref{c_pr_5},\ref{c_pr_10},\ref{c_pr_15} report the best curves out of 25 trials and the values of precision, recall, and reach which maximize the F1 score. Comparing these results to those in Table \ref{pr_300_best}, we see that the introduction of noisy pages somewhat decreases the effectiveness of the attack and preserves privacy.

\begin{table}[H]
    \begin{tabular}{|l|l|l|l|l|}
    \hline
    Attack     & F1 &Precision & Recall   & Reach      \\ \hline
    Pairwise   & 0.237559 &0.316000  & 0.957000 & 239 \\
    Neural Net & 0.418918 &0.765000  & 0.926000 & 19 \\ \hline
    \end{tabular}
    \caption{Attack performance with 5 page additions}
    \label{c_pr_5}
\end{table}
\vspace{-3mm}

\begin{table}[H]
    \begin{tabular}{|l|l|l|l|l|}
    \hline
    Attack     & F1 &Precision & Recall   & Reach      \\ \hline
    Pairwise   & 0.256262 &0.389000  & 0.751000 & 84 \\
    Neural Net & 0.486721 &0.981000  & 0.966000 & 3  \\ \hline
    \end{tabular}
    \caption{Attack performance with 10 page additions}
    \label{c_pr_10}
\end{table}
\vspace{-3mm}

\begin{table}[H]
    \begin{tabular}{|l|l|l|l|l|}
    \hline
    Attack     &F1 & Precision & Recall   & Reach      \\ \hline
    Pairwise   &0.292982 & 0.494000  & 0.720000 & 98 \\
    Neural Net & 0.413349 &0.763000  & 0.902000 & 15 \\ \hline
    \end{tabular}
        \caption{Attack performance with 15 page additions}
        \label{c_pr_15}
\end{table}

\section{Discussion}

Despite the coarseness of the semantic signals - categorical classifications for each page on a single website - our attack strategies were able to successfully identify some proportion of the sessions which were launched by the same user. This work establishes a lower bound on the potential consequences of any attack. A determined adversary could easily gain access to richer semantic signals. For example, they could use third party cookies to gather the pages visited by a user across multiple sites \cite{mayer2012third}. Alternatively, a first party website could log every action by a user on it is website and use these to construct site specific browsing fingerprints. In either case, the adversary would have access to a set of semantic signals significantly richer than those present in our data. In the best case for users, the adversary would merely replicated our performance. In the worst case, they might fare significantly better and identify more users. 

Additionally, though our pairwise scoring function and classifier achieved success, both strategies were relatively simple and straightforward. They demonstrated that the core problem of extracting and comparing browsing fingerprints could be successfully tackled through either supervised learning or pairwise scoring. More sophisticated techniques would likely greatly enhance the threat posed by such an attack. 

The poorer performance of the attacks on homepage partitioned sessions can be explained by the nature of homepage partitioning. While a visit to the homepage may occur if a user is starting a new session, it may also occur if a user was switching context within a single session. A user who started browsing about sports and then wanted to read more about news might return to the homepage to visit news articles. Partitioning on this homepage would then result in two very different sessions with little in common. The browsing fingerprints extracted for each would be unrepresentative of the user. 

Intuitively, it makes sense that the performance of an attack would decrease as the sample of users increases in size. At the granularity of data available, we expect a majority of users to be relatively indistinguishable. As our sample increases in size, the number of indistinguishable users will grow disproportionately, resulting in the attack misclassifying a greater number of session pairs. Interestingly, this seems to suggest that if an adversary wanted to identify a greater number of users, they should focus on smaller samples, ie. data from small websites. 

Though in our work we focused exclusively on web browsing, semantic identification attacks are applicable in a vastly greater scope. They may be used by a first party to gather usage analytics. The New York Times for example, may run a semantic identification attack on each of its registered accounts to see if multiple individuals are using the same account. For each session of browsing on a specific account, it could extract browsing fingerprint. If multiple unique fingerprints were extracted for a single account, it might deduce that multiple individuals were using this account. 

As expected, adding noise to the browsing sessions (i.e., spurious  pages visits) did reduce the effectiveness of the attack - though only slightly. Simply adding noise to a browsing session may not be sufficient to obscure the extracted fingerprint (especially if that noise is added uniformly to all sessions). Intuitively, we would likely need to obscure the most distinguishing parts of a session. This may require copying page visits from other sessions (not launched by  the original user) in order to confuse an adversary.

In summary, the results of our experiments demonstrate that an adversary employing our attacks could successfully extract browsing fingerprints from a subset of sessions and use those fingerprints to identify when two sessions were launched by the same user. We presented a framework for evaluating the success of these attacks in the context of an adversary's goals and demonstrate how different attacks could be advantageous to different adversaries.

\section{Acknowledgments}
We'd like to thank Ramakrishnan Srikant, Dan Boneh, Ramanathan Guha, and Mehran Sahami,  Lea Kissner, Scott Ellis, and Jonathan Mayer for their advice, and guidance on this project. 

\bibliographystyle{abbrv}
\bibliography{sources}


\end{document}